\documentclass[12pt,a4paper]{article}
\usepackage{graphicx}

\title{\bf BFKL predictions for inclusive three jet production at the LHC} 
\author{F. Caporale$^1$, G. Chachamis$^1$,  B. Murdaca$^2$, A. Sabio Vera$^{1}$\\ \\
{\small $^1$ Instituto de F{\' \i}sica Te{\' o}rica UAM/CSIC, Nicol{\'a}s Cabrera 15}\\ 
{\small \& Universidad Aut{\' o}noma de Madrid, E-28049 Madrid, Spain.}\\
{\small $^2$ Dipartimento di Fisica, Universit{\`a} della Calabria \&}\\
{\small Istituto Nazionale di Fisica Nucleare, Gruppo Collegato di Cosenza,}\\
{\small I-87036 Arcavacata di Rende, Cosenza, Italy.}
}

\begin{document} 

\maketitle 

\abstract

We define new observables sensitive to BFKL dynamics in the context of multijet production at the 
large hadron collider (LHC). We propose the study of the inclusive production of three jets well separated in rapidity from each other, with two of them being very forward. We show that the tagging of a third jet in the central region of rapidity allows for a very strong test of the BFKL formalism. In particular, we have studied two projections on azimuthal angles for the differential cross section which allow for the definition of many different observables whose behavior when varying the $p_t$ and  rapidity of the central jet is a distinct signal of BFKL dynamics. In order to reduce the theoretical uncertainties and influence of higher order corrections, we propose the study of ratios of correlation functions of products of cosines of azimuthal angle differences among the tagged jets.

\section{Introduction}

The large number of events already recorded and those to be produced in the near future at the Large Hadron Collider (LHC) offer a unique opportunity to disentangle the region of applicability of asymptotic calculations of scattering amplitudes in the high energy Regge limit. 

In this work we focus on the description of new observables where the Balitsky-Fadin-Kuraev-Lipatov (BFKL) formalism, at leading (LL) \cite{Lipatov:1985uk,Balitsky:1978ic,Kuraev:1977fs,Kuraev:1976ge,Lipatov:1976zz,Fadin:1975cb} and next-to-leading (NLL)~\cite{Fadin:1998py,Ciafaloni:1998gs} accuracy, should apply. In a nutshell, when calculating a scattering amplitude within this approach we single out those contributions at each order in perturbation theory with the largest numerical value when the center-of-mass energy, $\sqrt{s}$,  is asymptotically higher than any of the other Mandelstam invariants. These enhanced contributions are linked to the rapidity dependence of the observable under consideration.  From a phenomenological perspective, it is important to find the window of applicability for this formalism. This means to identify observables where the BFKL approach is distinct, {\it i.e.}, quantities where it fits the measured data and all other possible approaches (fixed order or other resummations implemented in general Monte Carlo event generators) fail. 

So far, the search for BFKL effects has had the general drawback of having collisions with too low $\sqrt{s}$ or rapidity differences among the tagged particles in the final state. A further problem has been to consider observables which are too inclusive as to be able to claim that the cross section under study could only be described by BFKL dynamics and nothing else. A canonical example is the growth of the hadron structure functions at low values of Bjorken $x$ in Deep Inelastic Scattering. Indeed it is possible to get a good fit of the combined HERA data for $F_{2,L}$ with a NLL BFKL calculation ({\it e.g.},~\cite{Hentschinski:2012kr,Hentschinski:2013id}).  However, it is equally possible to fit these data with other approaches. We must find other, less inclusive, observables to test small $x$ resummations. 

The LHC solves these problems since the available energies are much higher than at the Tevatron or HERA and there is enough statistics to allow for the study of very exclusive quantities, with strong kinematical cuts. The experimental challenges in this direction are remarkable since it is needed a large  rapidity span in the final states and a good resolution in azimuthal angles. We should not only consider  the usual ``growth with energy" signal, associated to the exchange of a hard pomeron, but also other footprints related to energy flow and azimuthal angle dependences. The latter are the main subject of the work here presented.

Typical BFKL observables at the LHC are the azimuthal angle ($\phi$) decorrelation of two tagged  forward jets widely separated in rapidity, $Y$, with associated inclusive mini-jet radiation, the so-called Mueller-Navelet jets~\cite{Mueller:1986ey}. This multiple emission  appears as a fast decrease of $\langle \cos{(n \, \phi)} \rangle$ as a function of $Y$~\cite{DelDuca:1993mn,Stirling:1994he,Orr:1997im,Kwiecinski:2001nh}. However, these differential distributions suffer from a large influence of collinear regions in phase space~\cite{Vera:2006un,Vera:2007kn}. This is due to the fact that $< \hspace{-.12cm}\cos{(n \phi)} \hspace{-.12cm}> \simeq \exp{(\alpha_s Y (\chi_n (1/2) - \chi_0(1/2)))}$, where $\alpha_s$ is the strong coupling and 
$\chi_n(\gamma)$ is, in Mellin space, the $n$-th Fourier component in $\phi$ of the BFKL kernel where the region $\gamma \simeq 1/2$ dominates for large $Y$. The $n=0$ component is very sensitive to collinear dynamics well  beyond the original multi-Regge kinematics. Even though it is possible to resum these collinear contributions ``on top" of the BFKL original calculation we believe that it is more important at present to fix the real region of applicability of the original BFKL formalism  by using observables which are far less sensitive to this collinear ``contamination". It is our target to find in this way distinct  BFKL observables. An important step in this direction was taken in~\cite{Vera:2006un,Vera:2007kn}  where it was proposed to remove the $n=0$ dependence by studying the ratios ${\cal C}_{m,n} = \langle \cos{(m \, \phi)} \rangle / \langle \cos{(n \, \phi)} \rangle$ which 
behave like $\sim \exp{(\alpha_s Y (\chi_m (1/2) - \chi_n (1/2)))}$. It is important to note that the BFKL kernel for $n \neq 0$ is insensitive to collinear regions, as it was shown in~\cite{Vera:2006un,Vera:2007kn}. It was also shown that these ${\cal C}_{m,n}$  ratios are very stable under radiative corrections, with the LO result (including running of the coupling) giving very similar results to the full NLL calculations. 

\begin{figure}
\vspace{-0.5cm}
\begin{center}
\includegraphics[height=2.5in]{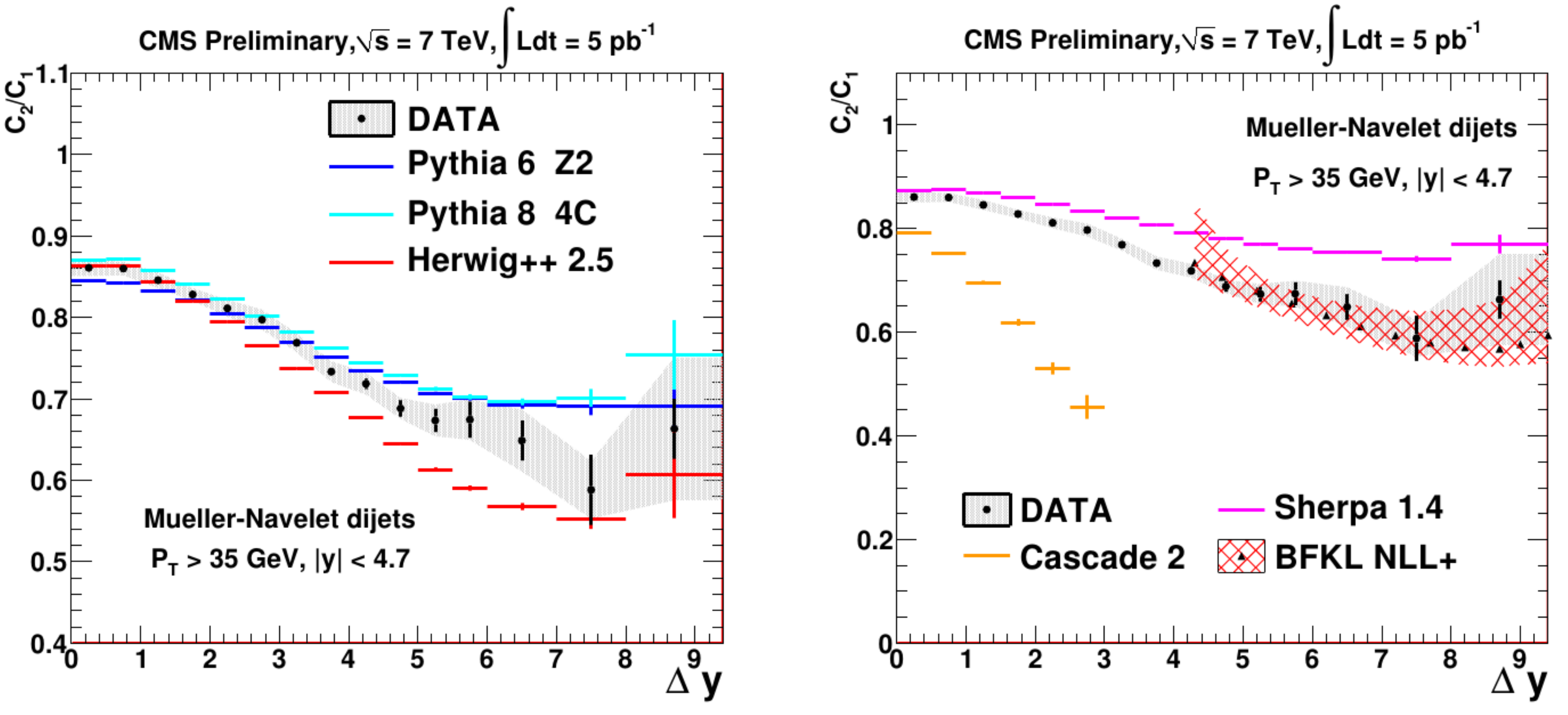}
\end{center}
\vspace{-0.6cm}
\caption{The ratio of average cosines ${\cal C}_{2,1} = C_2/C_1$ in bins of  $\Delta y = Y$, compared to various Monte Carlo models. Plots taken from~\cite{Ciesielski:2014dfa}.}
\label{C21vsMC}
\end{figure}

After the arrival of LHC data it has been seen that the NLL predictions, including NLO forward jet vertices, for the  ${\cal C}_{m,n}$ ratios are in agreement with the experimental results. Furthermore, these observables are so fine tuned  to the multi-Regge limit that it is difficult for other approaches to fit them with accuracy.
This can be seen in the recent studies presented in~\cite{Ducloue:2013bva,Caporale:2014gpa,Ciesielski:2014dfa}, see Fig.~\ref{C21vsMC}, where only a BFKL analysis at NLL is able to fit the large $Y$ tail of the Mueller-Navelet  ${\cal C}_{m,n}$  ratios  proposed in~\cite{Vera:2006un,Vera:2007kn}.

In our opinion, it is very important to continue along this line of work in the coming years of analysis of LHC data. In particular, it is needed to propose new quantities sensitive to BFKL dynamics, limiting the influence of the, otherwise widely dominant, collinear regions of phase space. In the coming Section we show that, if in the previously studied events with two tagged forward jets, we tag on a third, central, jet there will be many distinct new observables dominated by BFKL dynamics which are worth investigating both theoretically and experimentally. 
 
\section{Inclusive three jet production}

We propose to study events with two tagged forward jets, separated by a large rapidity span, and also tag on a third jet produced in the central region of rapidity, allowing for inclusive radiation in the remaining areas of the detectors.  In these processes it is possible to define many differential distributions in the transverse momentum, azimuthal angle and rapidity of the central jet, for fixed values of the four momenta of the forward jets. Our predictions for these observables will use the BFKL formalism to describe the inclusive multi jet emission taking place between the three tagged jets.

Before we proceed to explain the details of our calculation it is fair to highlight its limitations. We work at leading logarithmic accuracy, although we include running of the coupling effects. 
This will be improved in the future but we argue that, for distributions non sensitive to the zero conformal spin, higher order corrections will be small since the leading order prediction dominates the observables (the same argument was used in Ref.~\cite{Vera:2006un,Vera:2007kn} for the usual Mueller-Navelet jets case). We work with fixed four momenta for the two forward jets, this is for simplicity and clarity of presentation but a detailed extension including proper binning in these variables (which requires the introduction of collinear parton distributions) will be presented elsewhere together with further studies including the production of more than one jet in the central region of rapidity. Finally, although we provide useful analytic expressions for our observables, it is desirable to produce these results using Monte Carlo event generators, a work which is underway. 

The two tagged forward jets $A$ and $B$ have transverse momentum $\vec{k}_{A,B}$, azimuthal angle 
$\theta_{A,B}$ and rapidity $Y_{A,B}$. The central jet is characterized by $\vec{k}_J$, 
$\theta_J$ and $y_J$ and the differential cross section on these variables can be written in the form
\begin{eqnarray}
\frac{d^3 \sigma^{3-{\rm jet}}}{d^2 \vec{k}_J d y_J}   &=&  
 \frac{\bar{\alpha}_s }{\pi k_J^2}   \int d^2 \vec{p}_A \int d^2 \vec{p}_B \, 
 \delta^{(2)} \left(\vec{p}_A + \vec{k}_J- \vec{p}_B\right) \nonumber\\
&\times& \varphi \left(\vec{k}_A,\vec{p}_A,Y_A - y_J\right) 
 \varphi \left(\vec{p}_B,\vec{k}_B,y_J - Y_B\right),
 \label{Onejetemission}
\end{eqnarray}
where we assume that $Y_A > y_J > Y_B$ and $k_J$ lies above the experimental resolution 
scale. $\varphi$ are BFKL gluon Green functions normalized to $ \varphi \left(\vec{p},\vec{q},0\right) = \delta^{(2)} \left(\vec{p} - \vec{q}\right)$ and $\bar{\alpha}_s = \alpha_s N_c/\pi$.

In this paper we focus on quantities for which we find that the BFKL formalism will be both distinct from other approaches and very insensitive to higher order corrections. With this target in mind, we first integrate over the azimuthal angle of the central jet and over the difference in azimuthal angle between the two forward jets, $\Delta \phi \equiv \theta_A - \theta_B - \pi$, to define a quantity similar to the usual Mueller-Navelet case, {\it i.e.},  
\begin{eqnarray}
\int_0^{2 \pi} d \Delta \phi \, \cos{\left(M \Delta \phi \right)} 
\int_0^{2 \pi} d \theta_J\,\frac{d^3 \sigma^{3-{\rm jet}}}{d^2 \vec{k}_J d y_J}  &&  \\
&&\hspace{-7.4cm}  = \bar{\alpha}_s \sum_{L=0}^M 
 \int_{0}^\infty dp^2 \int_0^{2 \pi} d \theta 
 \frac{(-1)^M \left( \begin{array}{c}\hspace{-.2cm}M \\
\hspace{-.2cm}L\end{array} \hspace{-.18cm}\right)
 \left(k_J^2\right)^{\frac{L-1}{2}}  \, \left(p^2\right)^{\frac{M-L}{2}}
 \cos{\left(L \, \theta\right)}}{ \sqrt{\left(p^2 + k_J^2+ 2 \sqrt{p^2 k_J^2} \cos{\theta}\right)^{M}}}\nonumber\\
&&\hspace{-7.5cm} \times   \phi_{M} \left(p_A^2,p^2,Y_A-y_J\right)
\phi_{M} \left(p^2+ k_J^2 + 2 \sqrt{p^2 k_J^2} \cos{\theta},p_B^2,y_J-Y_B\right),
\nonumber
\end{eqnarray}
where ($\psi$ is the logarithmic derivative of Euler's gamma function)
\begin{eqnarray}
\phi_{n} \left(p_A^2,p_B^2,Y\right)  &=&  
2 \int_0^\infty d \nu   
\cos{\left(\nu \ln{\frac{p_A^2}{p_B^2}}\right)}  \frac{e^{\bar{\alpha}_s  \chi_{|n|} \left(\nu\right) Y}}{\pi \sqrt{p_A^2 p_B^2} }, \label{phin}\\
 \chi_{n} \left(\nu\right) &=& 2 \psi (1) - \psi \left( \frac{1+n}{2} + i \nu\right) - \psi \left(\frac{1+n}{2} - i \nu\right).
\end{eqnarray}
One of the experimental observables we want to highlight here corresponds to the mean 
value of the cosine of $\Delta \phi$ in the recorded events:
\begin{eqnarray}
\langle \cos{\left(M \left(\theta_A - \theta_B - \pi\right) \right)} \rangle = 
\frac{\int_0^{2 \pi} d \Delta \phi \, \cos{\left(M \Delta \phi \right)} 
\int_0^{2 \pi} d \theta_J\, \frac{d^3 \sigma^{3-{\rm jet}}}{d^2 \vec{k}_J d y_J} }{\int_0^{2 \pi} d \Delta \phi 
\int_0^{2 \pi} d \theta_J\,\frac{d^3 \sigma^{3-{\rm jet}}}{d^2 \vec{k}_J d y_J}  }.
\end{eqnarray}
As we have already mentioned, the perturbative stability (including renormalization scale dependence) of our predictions is much better (see~\cite{Caporale:2013uva} for a related discussion) if we remove the contribution from the zero conformal spin which corresponds to the index $n=0$ in Eq.~(\ref{phin}). We can achieve this by defining the ratios
\begin{eqnarray}
{\cal R}_{N}^M &=& \frac{\langle \cos{\left(M \left(\theta_A - \theta_B - \pi\right) \right)} \rangle}{\langle \cos{\left(N \left(\theta_A - \theta_B - \pi\right) \right)} \rangle},
\label{RMN}
\end{eqnarray}
where we consider $M,N$ as positive integers. 

\begin{figure}
\vspace{-0.5cm}
\begin{center}
\includegraphics[height=3.in]{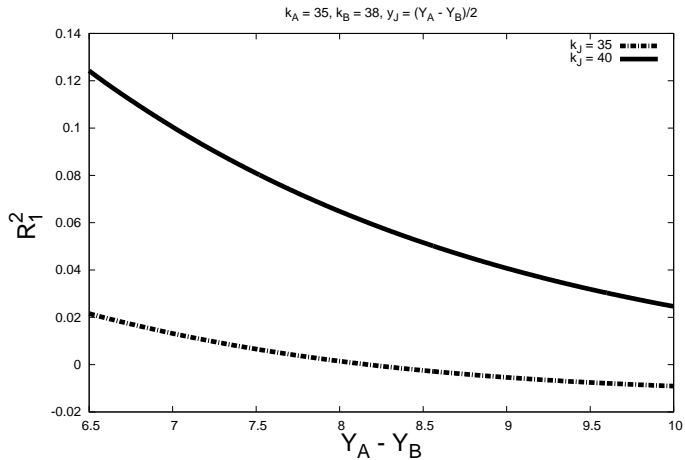}
\end{center}
\vspace{-1.4cm}
\caption{A study of the ratio ${\cal R}_{1}^2$ as defined in Eq.~(\ref{RMN}) for fixed values of the $p_t$ of the two forward jets and two values of the $p_t$ of the tagged central jet, as a function of the rapidity difference between the two forward jets for the rapidity of the central jet chosen as $y_J = (Y_A - Y_B)/2$.}
\label{R21}
\end{figure}

Many studies can be performed with these ratios but here he offer a first one where we fix the transverse momenta of the forward jets to 
$k_A=35$ GeV and $k_B = 38$ GeV. We also fix the rapidity of the central jet to be one half of the rapidity difference between the two forward jets: $y_J = (Y_A - Y_B)/2$ simply because this allows us to connect with the well-known Mueller-Navelet jets. In this way we can plot, {\it e.g.}, the ratio ${\cal R}_{1}^2$ in Fig.~\ref{R21} for two values of the transverse momentum of the central jet $k_J = 35, 40$ GeV where we can see that this ratio decreases as a function of $Y_A-Y_B$. This is a consequence of having an increase in the available phase space for inclusive minijet radiation and that the $n=1$ component decreases which energy slower that the $n=2$ contribution. 

In the BFKL formalism we have that the larger the $n$ the slower the evolution with rapidity differencies. This is very important since it is distinct from other approaches where QCD coherence is introduced as it was shown in Ref.~\cite{Chachamis:2011rw}. We believe this is the reason why the usual Monte Carlo event generators fail to properly describe the Mueller-Navelet ratios proposed in~\cite{Vera:2006un,Vera:2007kn} and will probably also fail to describe the ones here investigated. 

In the last part of this work, we want to propose new observables whose associated distributions have a very different behavior to the ones characteristic of the Mueller-Navelet case. These new distributions are defined using the projections on the two relative azimuthal angles formed by each of the forward jets with the central jet, $\theta_A - \theta_J - \pi$ and $\theta_J - \theta_B - \pi$, in the form
\begin{eqnarray}
\int_0^{2 \pi} d \theta_A \int_0^{2 \pi} d \theta_B \int_0^{2 \pi} d \theta_J \cos{\left(M \left( \theta_A - \theta_J - \pi\right)\right)} && \nonumber\\
&&\hspace{-5.cm}\cos{\left(N \left( \theta_J - \theta_B - \pi\right)\right)}
\frac{d^3 \sigma^{3-{\rm jet}}}{d^2 \vec{k}_J d y_J}  \nonumber\\
&&\hspace{-8.5cm}= \bar{\alpha}_s \sum_{L=0}^{N} 
\left( \begin{array}{c}
\hspace{-.2cm}N \\
\hspace{-.2cm}L\end{array} \hspace{-.18cm}\right)
 \left(k_J^2\right)^{\frac{L-1}{2}} 
\int_{0}^\infty d p^2 \, \left(p^2\right)^{\frac{N-L}{2}} \\
&&\hspace{-6.cm}\int_0^{2 \pi}  d \theta    \frac
{   (-1)^{M+N} \cos{ \left(M \theta\right)} \cos{\left((N-L) \theta\right)}
}{
 \sqrt{\left(p^2 + k_J^2+ 2 \sqrt{p^2 k_J^2} \cos{\theta}\right)^{N}}
}
\nonumber\\
&&\hspace{-8.5cm}  \times \, \, \phi_{M} \left(p_A^2,p^2,Y_A-y_J\right)
\phi_{N} \left(p^2+ k_J^2 + 2 \sqrt{p^2 k_J^2}\cos{\theta},p_B^2,y_J-Y_B\right).\nonumber
\end{eqnarray}

The experimentally relevant observable is the mean value in the selected events of the two cosines, {\it i.e.}
\begin{eqnarray}
\langle \cos{\left(M \left( \theta_A - \theta_J - \pi\right)\right)}  
\cos{\left(N \left( \theta_J - \theta_B - \pi\right)\right)}
\rangle && \\
&&\hspace{-8.5cm} = \frac{\int_0^{2 \pi} d \theta_A d \theta_B d \theta_J \cos{\left(M \left( \theta_A - \theta_J - \pi\right)\right)}  \cos{\left(N \left( \theta_J - \theta_B - \pi\right)\right)}
\frac{d^3 \sigma^{3-{\rm jet}}}{d^2 \vec{k}_J d y_J} }{\int_0^{2 \pi} d \theta_A d \theta_B d \theta_J 
\frac{d^3 \sigma^{3-{\rm jet}}}{d^2 \vec{k}_J d y_J} }.\nonumber
\end{eqnarray}
As before, in order to have optimal perturbative convergence and eliminate collinear contamination, we can remove the contributions from zero conformal spin by defining the ratios:
\begin{eqnarray}
{\cal R}_{P,Q}^{M,N} &=& \frac{\langle \cos{\left(M \left( \theta_A - \theta_J - \pi\right)\right)}  
\cos{\left(N \left( \theta_J - \theta_B - \pi\right)\right)}
\rangle}{\langle \cos{\left(P \left( \theta_A - \theta_J - \pi\right)\right)}  
\cos{\left(Q \left( \theta_J - \theta_B - \pi\right)\right)}
\rangle}
\label{RMNPQ}
\end{eqnarray}
and consider $M,N,P,Q >0$ as integer numbers. 

\begin{figure}
\vspace{-0.5cm}
\begin{center}
\includegraphics[height=3.in]{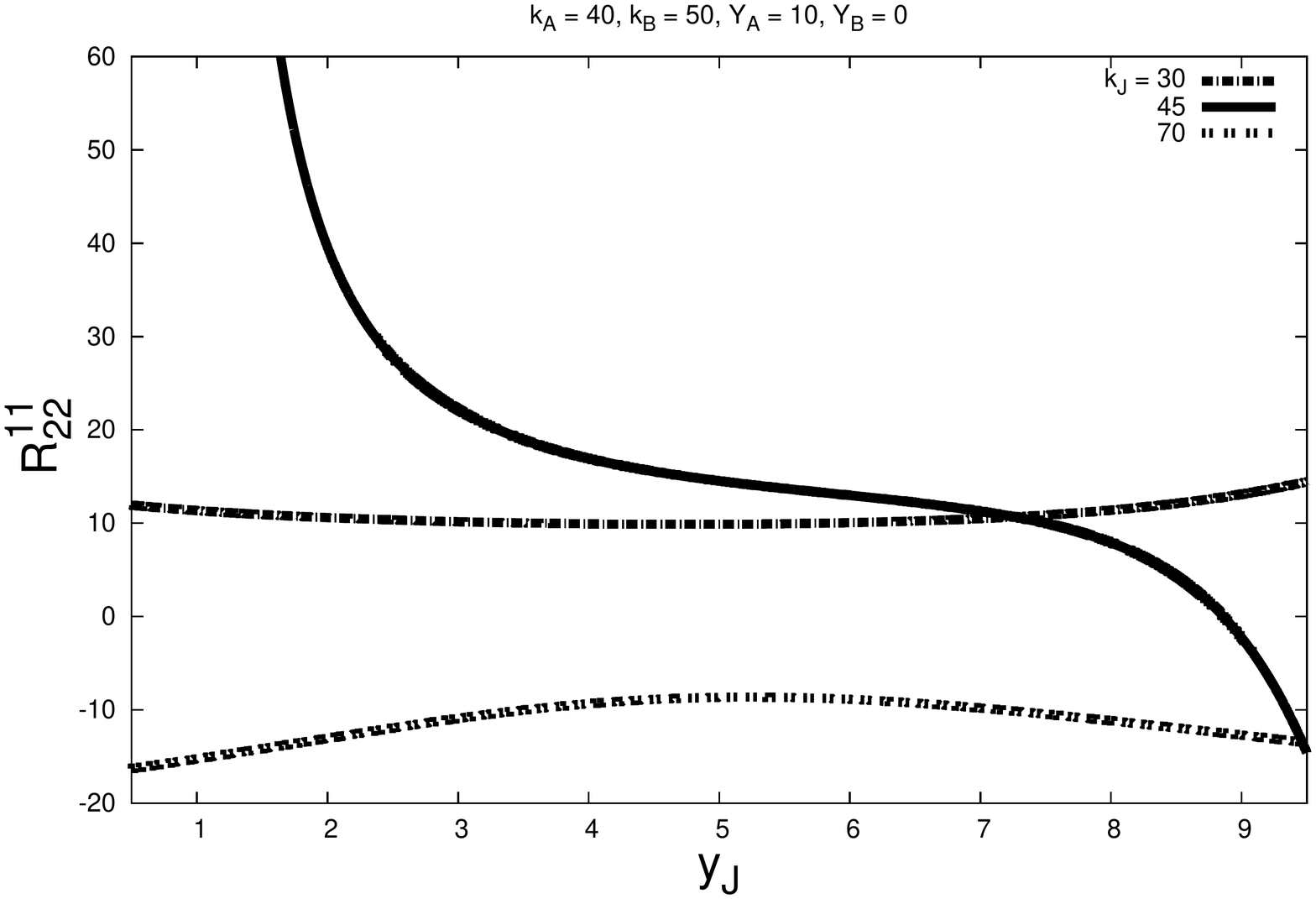}\\
\vspace{-1.4cm}
\includegraphics[height=3.in]{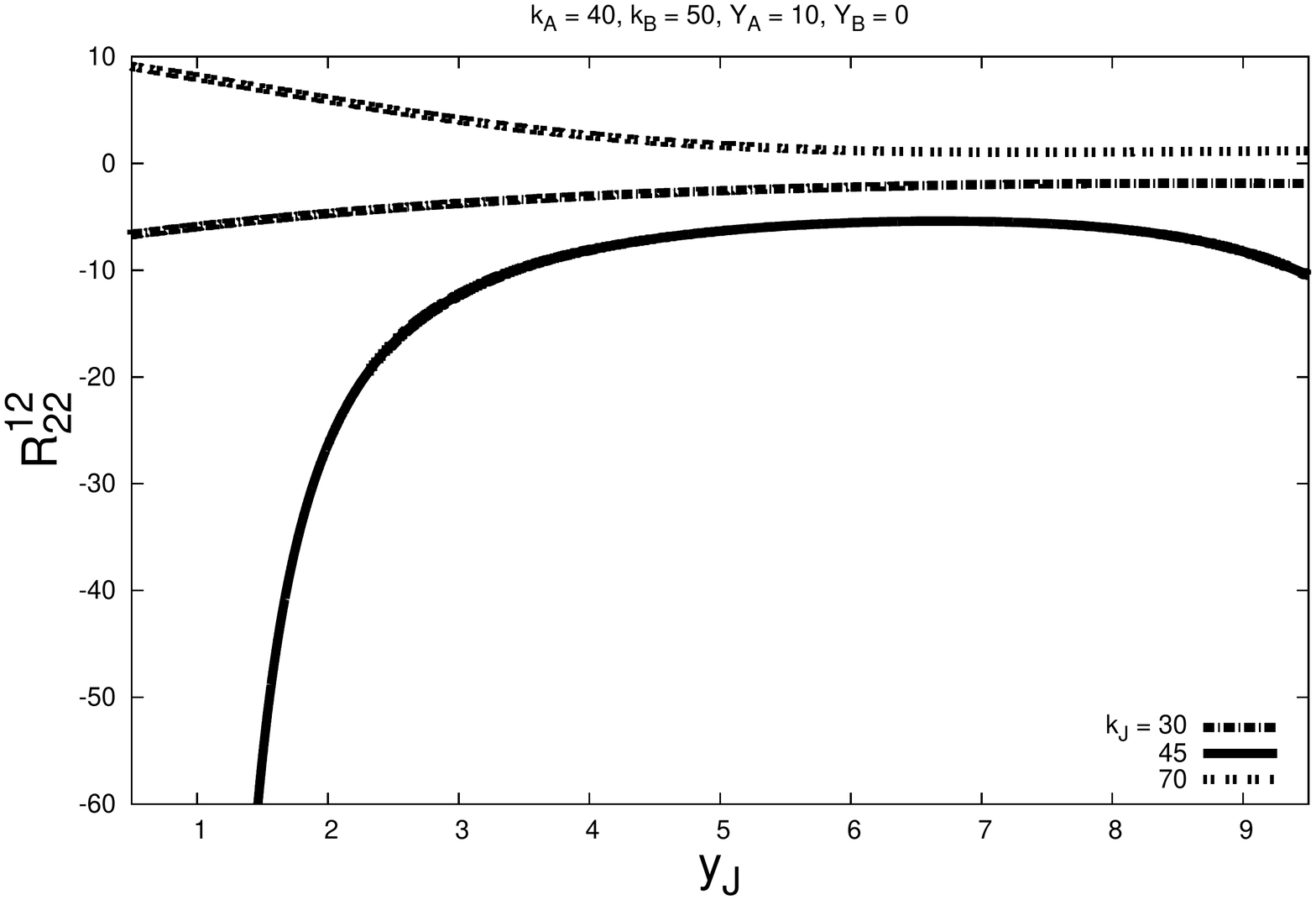}\\
\vspace{-1.4cm}
\includegraphics[height=3.in]{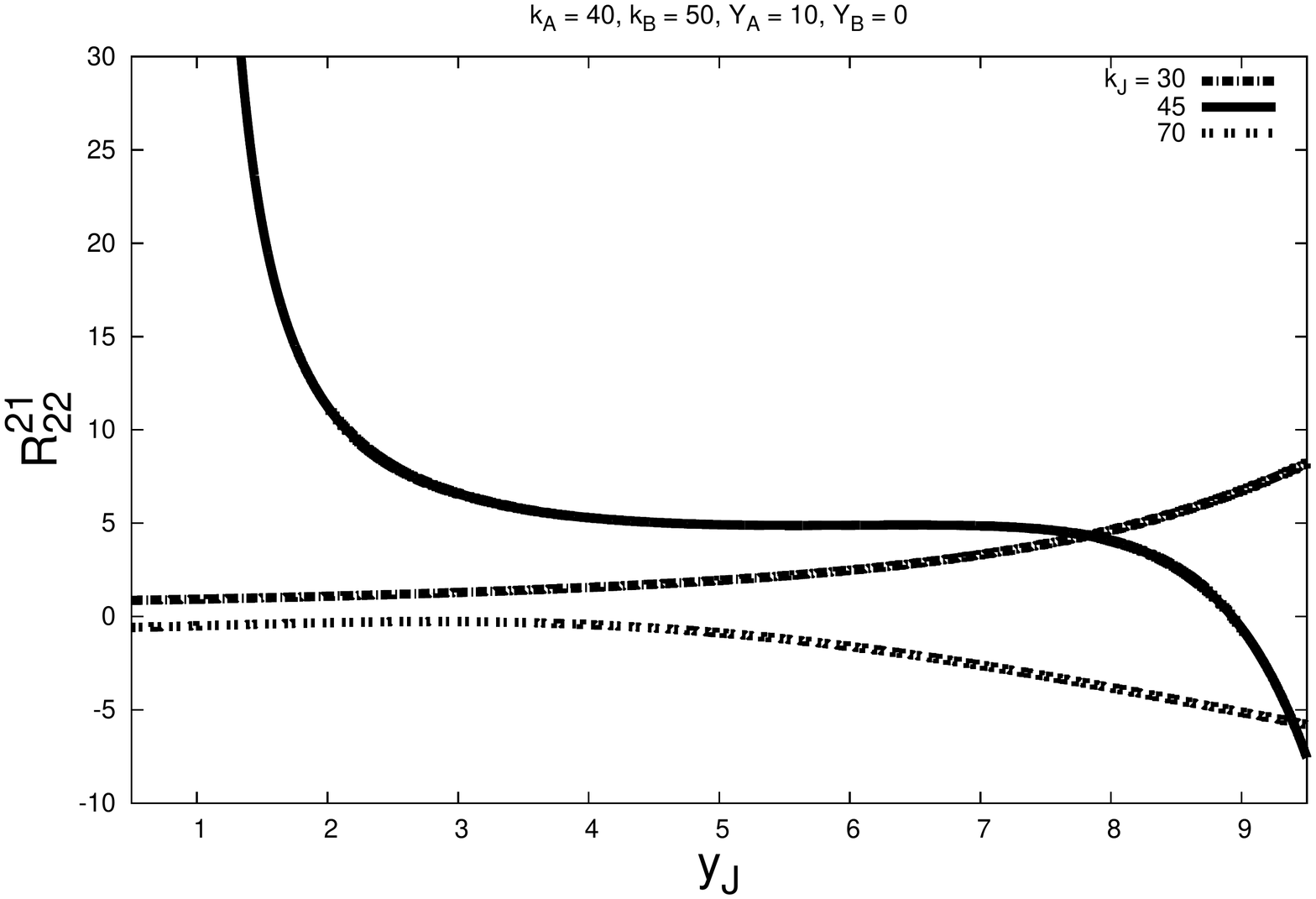}
\end{center}
\vspace{-1.4cm}
\caption{A study of the ratios ${\cal R}_{2,2}^{1,1}$,  ${\cal R}_{2,2}^{1,2}$ and ${\cal R}_{2,2}^{2,1}$ as defined in Eq.~(\ref{RMNPQ}) for fixed values of the $p_t$ of the two forward jets and three values of the $p_t$ of the tagged central jet, as a function of the rapidity of the central jet $y_J$.}
\label{RMNPQfigure}
\end{figure}

We can now investigate many momenta configurations. As an example here we show some ratios ${\cal R}_{P,Q}^{M,N}$ with $M,N=1,2$ fixing the momenta of the forward jets to $k_A=40$ GeV and $k_B=50$ GeV and 
their rapidities to $Y_A=10$ and $Y_B=0$. For the transverse momentum of the central jet we choose three values $k_J= 30, 45, 70$ GeV and we vary the rapidity of the central jet $y_J$ in between the two rapidities of the forward jets. We show the results in Fig.~\ref{RMNPQfigure}. 
These distributions are proving the fine structure of the QCD radiation in the high energy limit. They gauge the relative weights of each conformal spin contribution to the total cross section. We expect the LHC data to agree with these results specially in the regions where 
$y_J$ is closer to $(Y_A - Y_B)/2$. It will be very interesting to see how they compare to the predictions from fixed order analysis and Monte Carlo event generators.

\section{Summary \& Outlook}

We have addressed the problem of defining new observables which can be sensitive to BFKL dynamics in the context of the LHC physics program. We have focussed on the inclusive production of three jets well separated in rapidity from each other. The BFKL resummation contains much more information than just the growth of cross sections: when projected on azimuthal angles it is expressed as a sum of infinite components from which only the first one ($n=0$ Fourier component) grows with energy, all the others decrease. This fact has already been used to discriminate between this type of resummations and other approaches in the case of Mueller-Navelet jets, with two tagged jets in the very forward directions. Here we have proposed to tag a third jet in the central region of rapidity allowing us to test the BFKL formalism at a more exclusive level. This third jet connects with the two forward ones via two gluon Green functions and the further in rapidity it is emitted from them, the more decorrelated in azimuthal angles they will be. 

The projection on azimuthal angles we propose allows for the definition of many different observables whose behavior when varying the $p_t$ and  rapidity of the central jet is a distinct signal of BFKL dynamics. In order the reduce the theoretical uncertainties we have put forward the study of ratios of correlation functions of products of cosines of azimuthal angle differences among the tagged jets in Eqs.~(\ref{RMN}) and~(\ref{RMNPQ}). This suppresses the collinear dynamics and reduces the influence of the parton distribution functions. 

This program is important since there are uncertainties in the BFKL approach itself which need to be fixed and the current data recorded at the LHC will be crucial to do so. Only a fair comparison to experimental data can solve many of these theoretical questions. We believe the type of observables proposed in this work will be crucial to define the region of phenomenological applicability of the BFKL resummation.

\begin{flushleft}
{\bf \large Acknowledgements}
\end{flushleft}
G.C. acknowledges support from the MICINN, Spain, under contract FPA2013-44773-P. 
A.S.V. acknowledges support from Spanish Government (MICINN (FPA2010-17747,FPA2012-32828)) and, together with F.C and B.M., to the Spanish MINECO Centro de Excelencia Severo Ochoa Programme (SEV-2012-0249). The work of B.M. was supported in part by the grant RFBR-13-02-90907 and by the European
Commission, European Social Fund and Calabria Region, that disclaim any liability for the use
that can be done of the information provided in this paper.


\begin{thebibliography}{10}

  
\bibitem{Lipatov:1985uk}
  L.~N.~Lipatov,
  Sov.\ Phys.\ JETP {\bf 63} (1986) 904
   [Zh.\ Eksp.\ Teor.\ Fiz.\  {\bf 90} (1986) 1536].
  
\bibitem{Balitsky:1978ic}
  I.~I.~Balitsky and L.~N.~Lipatov,
  Sov.\ J.\ Nucl.\ Phys.\  {\bf 28} (1978) 822
   [Yad.\ Fiz.\  {\bf 28} (1978) 1597].
  
\bibitem{Kuraev:1977fs}
  E.~A.~Kuraev, L.~N.~Lipatov and V.~S.~Fadin,
  Sov.\ Phys.\ JETP {\bf 45} (1977) 199
   [Zh.\ Eksp.\ Teor.\ Fiz.\  {\bf 72} (1977) 377].
  
\bibitem{Kuraev:1976ge}
  E.~A.~Kuraev, L.~N.~Lipatov and V.~S.~Fadin,
  Sov.\ Phys.\ JETP {\bf 44} (1976) 443
   [Zh.\ Eksp.\ Teor.\ Fiz.\  {\bf 71} (1976) 840]
   [Erratum-ibid.\  {\bf 45} (1977) 199].
  
\bibitem{Lipatov:1976zz}
  L.~N.~Lipatov,
  Sov.\ J.\ Nucl.\ Phys.\  {\bf 23} (1976) 338
   [Yad.\ Fiz.\  {\bf 23} (1976) 642].
  
\bibitem{Fadin:1975cb}
  V.~S.~Fadin, E.~A.~Kuraev and L.~N.~Lipatov,
  Phys.\ Lett.\ B {\bf 60} (1975) 50.
  
\bibitem{Fadin:1998py}
  V.~S.~Fadin and L.~N.~Lipatov,
  Phys.\ Lett.\ B {\bf 429} (1998) 127
  [hep-ph/9802290].
  
\bibitem{Ciafaloni:1998gs}
  M.~Ciafaloni and G.~Camici,
  Phys.\ Lett.\ B {\bf 430} (1998) 349
  [hep-ph/9803389].


\bibitem{Hentschinski:2012kr}
  M.~Hentschinski, A.~Sabio Vera and C.~Salas,
  Phys.\ Rev.\ Lett.\  {\bf 110} (2013) 041601
  [arXiv:1209.1353 [hep-ph]].
  
\bibitem{Hentschinski:2013id}
  M.~Hentschinski, A.~Sabio Vera and C.~Salas,
  Phys.\ Rev.\ D {\bf 87} (2013) 076005
  [arXiv:1301.5283 [hep-ph]].
  
  
\bibitem{Mueller:1986ey}
  A.~H.~Mueller and H.~Navelet,
  Nucl.\ Phys.\ B {\bf 282} (1987) 727.
  
  
\bibitem{DelDuca:1993mn}
  V.~Del Duca and C.~R.~Schmidt,
  Phys.\ Rev.\ D {\bf 49} (1994) 4510
  [hep-ph/9311290].
  
  
\bibitem{Stirling:1994he}
  W.~J.~Stirling,
  Nucl.\ Phys.\ B {\bf 423} (1994) 56
  [hep-ph/9401266].
  
  
\bibitem{Orr:1997im}
  L.~H.~Orr and W.~J.~Stirling,
  Phys.\ Rev.\ D {\bf 56} (1997) 5875
  [hep-ph/9706529].
  
  
\bibitem{Kwiecinski:2001nh}
  J.~Kwiecinski, A.~D.~Martin, L.~Motyka and J.~Outhwaite,
  Phys.\ Lett.\ B {\bf 514} (2001) 355
  [hep-ph/0105039].
  
\bibitem{Vera:2006un}
  A.~Sabio Vera,
  Nucl.\ Phys.\ B {\bf 746} (2006) 1
  [hep-ph/0602250].
  
\bibitem{Vera:2007kn}
  A.~Sabio Vera and F.~Schwennsen,
  Nucl.\ Phys.\ B {\bf 776} (2007) 170
  [hep-ph/0702158 [HEP-PH]].
  
\bibitem{Ducloue:2013bva}
  B.~Ducloue, L.~Szymanowski and S.~Wallon,
  Phys.\ Rev.\ Lett.\  {\bf 112} (2014) 082003
  [arXiv:1309.3229 [hep-ph]].

\bibitem{Caporale:2014gpa}
  F.~Caporale, D.~Y.~Ivanov, B.~Murdaca and A.~Papa,
  Eur.\ Phys.\ J.\ C {\bf 74} (2014) 3084
  [arXiv:1407.8431 [hep-ph]].
  
\bibitem{Ciesielski:2014dfa}
  R.~Ciesielski,
  arXiv:1409.5473 [hep-ex].
  
  
  
\bibitem{Caporale:2013uva}
  F.~Caporale, B.~Murdaca, A.~Sabio Vera and C.~Salas,
  Nucl.\ Phys.\ B {\bf 875} (2013) 134
  [arXiv:1305.4620 [hep-ph]].
  

\bibitem{Chachamis:2011rw}
  G.~Chachamis, M.~Deak, A.~Sabio~Vera and P.~Stephens,
  Nucl.\ Phys.\ B {\bf 849} (2011) 28
  [arXiv:1102.1890 [hep-ph]].


\end{thebibliography}
\end{document}